\documentclass[10pt]{iopart}

\usepackage{graphicx,latexsym,color}
\usepackage{cite}

\usepackage{amsgen,amsfonts,amsbsy,amssymb}

\usepackage{listings}

\usepackage{bm}
\usepackage{tikz}
\usepackage{pgfplots}

\newcommand{\ie}{\textit{i.e.}\/, }
\newcommand{\eg}{\textit{e.g.}\/, }
\newcommand{\cf}{\textit{cf.}\/, }

\providecommand*{\mrm}[1]{\mathrm{#1}}
\providecommand*{\unit}[1]{\ensuremath{\mrm{\,#1}}}
\providecommand*{\eu}{\ensuremath{\mrm{e}}}

\providecommand*{\ju}{\ensuremath{\mrm{j}}}

\newcommand{\minimize}{\mrm{minimize}}

\def\XXint#1#2#3{{\setbox0=\hbox{$#1{#2#3}{\int}$}
     \vcenter{\hbox{$#2#3$}}\kern-.5\wd0}}

\begin{document}

\title{A parametric model for the changes in the complex valued conductivity of a lung during tidal breathing}

\author{Sven~Nordebo$^1$, Mariana~Dalarsson$^1$, Davood~Khodadad$^1$, Beat~M\"{u}ller$^2$, Andreas~Waldman$^2$, Tobias~Becher$^3$, Inez~Frerichs$^3$,
Louiza~Sophocleous$^4$, Daniel~Sj\"{o}berg$^5$, Nima~Seifnaraghi$^6$, Richard~Bayford$^6$}

\address{$^1$ Department of Physics and Electrical Engineering, Linn\ae us University, 351 95 V\"{a}xj\"{o}, Sweden.
E-mail: \{sven.nordebo,mariana.dalarsson,davood.khodadad\}@lnu.se.}
\address{ $^2$ Swisstom AG, Schulstrasse 1, CH-7302 Landquart, Switzerland. E-mail: \{bmu,awa\}@swisstom.com}
\address{$^3$ Department of Anaesthesiology and Intensive Care Medicine, University Medical Centre Schleswig-Holstein, Campus Kiel, 24105 Kiel, Germany. 
E-mail: \{Tobias.Becher,Inez.Frerichs\}@uksh.de.}
\address{$^4$ The KIOS Research Center, Department of Electrical and Computer Engineering, University of Cyprus, Nicosia, 
Cyprus. E-mail: louiza.sophocleous@gmail.com.}
\address{$^5$ Department of Electrical and Information Technology, Lund University, Box 118, 
221 00 Lund, Sweden. E-mail: daniel.sjoberg@eit.lth.se.}
\address{$^6$ Department of Natural Sciences, Middlesex University, Hendon campus, 
The Burroughs, London, NW4 4BT, United Kingdom. E-mail: nima.seifnaraghi@gmail.com, R.Bayford@mdx.ac.uk.}

\maketitle
\ioptwocol

\begin{abstract}
Classical homogenization theory based on the Hashin-Shtrikman coated ellipsoids is used to model the changes in the complex valued conductivity (or admittivity)
of a lung during tidal breathing. Here, the lung is modeled as a two-phase composite material where the alveolar air-filling corresponds to the inclusion phase.
The theory predicts a linear relationship between the real and the imaginary parts of the change in the complex valued conductivity of a lung during tidal breathing,
and where the loss cotangent of the change is approximately the same as of the effective background conductivity and hence easy to estimate.
The theory is illustrated with numerical examples, as well as by using reconstructed Electrical Impedance Tomography (EIT) images 
based on clinical data from an ongoing study within the EU-funded CRADL project. 
The theory may be potentially useful for improving the imaging algorithms and clinical evaluations in connection with lung EIT for 
respiratory management and monitoring in neonatal intensive care units.
\end{abstract}

\section{Introduction}\label{sect:introduction}

Electrical Impedance Tomography (EIT) is a non-radiative, inexpensive technique that can facilitate real
time dynamic monitoring of regional lung aeration and ventilation for clinical use \cite{Frerichs2000}. 
The approach lacks spatial resolution, but it benefits largely from its high temporal resolution and 
is therefore currently emerging as a technique that can potentially reduce complications and disability in preterm babies by continuous bedside monitoring
and respiratory management \cite{Carlisle2010,Frerichs+etal2017}.

From a mathematical/physical point of view, EIT constitutes an ill-posed inverse problem \cite{Somersalo+Cheney+Isaacson1992,Cheney+etal1999,Bayford2006,Nordebo+etal2010b},
and the use of EIT as a successful imaging modality relies therefore on the effectiveness of creating difference conductivity images rather than to generate
absolute reconstructions, see \eg \cite{Adler+Guardo1996,Adler+etal2009}. The difference imaging approach benefits from the linearization of the forward
problem, and it alleviates much of the sensitivity to sensor imperfections as well as the unknown background parameter values.
In lung EIT,  tidal images can be created by using a breath detector \cite{Khodadad+etal2017}  
providing difference voltage data that is perfectly synchronized with the time of end-expiration and end-inspiration, defining the breathing cycle. 
For these tidal images, EIT related lung function parameters such as Center of Ventilation, Silent Spaces and ventilation distribution, etc., can then be calculated \cite{Frerichs+etal2017}.

Early EIT systems were usually designed for operation at very low frequencies, typically in the kilohertz range \cite{Cheney+Isaacson1995a,Adler+etal1997},
where the conductivity of biological tissue usually is considered to be purely resistive (even though this is not entirely true \cite{Gabriel+etal1996b}).
However, the need of improved image quality in both the spatial and the temporal domains is nowadays driving the development of the commercial 
EIT systems and their data acquisition hardware to higher speeds (typically in the range of several hundreds of kilohertz) and hence their performance requirements
imply that the imaginary part of the complex valued conductivity (or admittivity) of human tissue no longer can be disregarded. 
Notably, this constitutes a technical challenge, but it can also be a great asset due to the additional clinical information carried by the permittivity of the tissue.
To this end, there is no fundamental limitation associated with the reconstruction of the complex valued conductivity using standard optimization algorithms such as in
 \cite{Somersalo+Cheney+Isaacson1992,Cheney+Isaacson1995a,Cheney+etal1999,Adler+etal2009,Nordebo+etal2010b}, 
nor with the new developments such as the D-bar methods \cite{Isaacson+etal2006,Hamilton+etal2017}.
In this paper, we investigate a mathematical/physical mechanism that can model the changes in the complex valued
conductivity of a lung during tidal breathing. This is particularly interesting as it is already empirically known that the EIT pixel sum of (real valued) conductivity changes
within a particular region of interest is almost linearly related to the changes in lung volume, see \eg \cite{Adler+etal1997}.

A comprehensive overview on the dielectric spectral properties of biological tissue, including modeling, measurements and literature is
given in \cite{Gabriel+etal1996a,Gabriel+etal1996b,Gabriel+etal1996c}. 
In particular, as a realistic estimate of the background conductivity of an inflated lung we have used here the real valued conductivity and permittivity
data from \cite[Fig.~2e on p.~2257]{Gabriel+etal1996b} which is of bovine origin.
Measurement data from human tissue, and in particular from neonatal patients is obviously very difficult to obtain.
Hence, in order to estimate the typical volume fraction of air during tidal breathing we have used data regarding the lung volume of an adult male 
and the corresponding condensed matter weight from \cite{Ganong2003} and \cite{Molina+DiMaio2012}, respectively.

A basic model predicting the dielectric properties of lung tissue as a function of air content was given in  \cite{Nopp+etal1993}, 
and which has been followed by several other works, see \eg \cite{Nopp+etal1997} and \cite{Wang2014} with references.
The basic work in \cite{Nopp+etal1993} was followed by a comprehensive study based on EIT spectroscopy measurements
and an improved model taking both the air content and tissue dispersion into account \cite{Nopp+etal1997}.
In \cite{Wang2014} is given an experimental study of dielectric properties of human lung tissue in vitro and
its dependency on the air filling factor. Human lung tissue specimens from more than 100 patients were investigated with respect to the differences
in the impedance spectrum for cancerous and normal tissue, and where the cancerous tissue typically show
much larger conductivity values due to the deterioration of the alveolar structure. The same can also be said about
interstitial pneumonia \cite{Nopp+etal1993}, where the increased conductivity is explained by the increased
thickness of the alveolar walls. Notably, the investigations on the dielectric properties of lung tissue as a function of air content
in \cite{Nopp+etal1993} and \cite{Wang2014} are {\em in vitro}, with volume fractions of air up to about 58-60\unit{\%}, whereas
our estimates of the air content with reference to \cite{Ganong2003} and \cite{Molina+DiMaio2012} are rather in the range 75-78\unit{\%} for tidal breathing {\em in vivo}.

The dielectric measurements in \cite{Nopp+etal1993} were made on the
excised lungs of slaughtered calves in the frequency range of 5 \unit{kHz} to 100 \unit{kHz},
and a simple theory was developed to model the complex valued conductivity as a function of air filling.
The theoretical model in \cite{Nopp+etal1993} is based on the deformation of a cube-shaped alveolus where the conductivity is given approximately
as the conductivity of the enclosing walls of a square cylinder where the volume of the (periodic) wall system is kept fixed
and the cube size and the wall thickness are variable to account for the deformation of the epithelial cells and blood vessels through the expansion of the alveoli.
The parameters of the model are adjusted by histological investigations.
In this way, the thinning of the alveolar walls can explain the decrease of effective conductivity and permittivity of the lung 
as a function of an increased air filling.

In this paper, we derive an alternative physical model based on classical homogenization theory \cite{Sihvola1999,Milton2002} to predict the changes in the complex valued 
conductivity of a lung during tidal breathing. The lung is modeled here as a two-phase composite material where the alveolar air-filling corresponds to the inclusion phase 
and the exterior phase is due to the blood and tissue.
The parametric model is based on classical Hashin-Shtrikman/Maxwell-Garnett theory \cite{Sihvola1999,Milton2002} and is simple
and well-suited for an analytical study on the changes in the effective conductivity of the lung due to the corresponding small changes in the volume fraction of air during tidal breathing. This model is in many ways similar to that of \cite{Nopp+etal1993} (an effective conductivity model with thinning of the alveolar walls, etc.),
but it has a rigorous foundation in homogenization theory. In particular, the parametric model based on the Hashin-Shtrikman (HS) assemblage \cite{Milton2002}
comprises a fully three-dimensional homogenization of the lung tissue with ellipsoidal shaped alveoli constituting the inclusion phase.
The HS assemblage furthermore accounts for the random nature of the alveolar structure, as
seen in \eg \cite[Figs.~14 and 15 on p.~711]{Nopp+etal1993}, by the inherent variation of the HS scaling and positioning of the prototype ellipsoid.
Hence, it may be argued from the figures in \cite{Nopp+etal1993}  that the shape of the alveoli is typically more round than square and that
the close packing of alveoli at high air content can be achieved due to a large variation in their sizes, similar to the HS assemblage.

The presented parametric model predicts an almost linear relationship between the real and the imaginary parts of the changes in the complex valued
conductivity of a lung during tidal breathing, and where the loss cotangent of the change is approximately the same as of the effective background conductivity 
and hence easy to estimate.
It is expected that this a priori knowledge can be useful in the development of new improved image reconstruction
algorithms exploiting complex valued measurement data, and/or to define new clinically useful outcome parameters in lung EIT.
The theoretical study is based on realistic parameter choices for the conductivity of an inflated lung \cite{Gabriel+etal1996b},
and hence the corresponding loss cotangent is estimated to be in the order of about $\cot\delta=0.2$ at 200\unit{kHz}. 
It should be noted that, even at low frequencies, the loss cotangent of the inflated lung may not be negligible (due to the very high relative permittivity of tissue at low frequencies),
and is expected to be in the order of about $\cot\delta=0.1$ at 1\unit{kHz}. 
At higher frequencies, it will increase to about $\cot\delta=0.3$ at 1\unit{MHz} \cite{Gabriel+etal1996b}.

The theoretical study is illustrated with numerical examples, as well as with clinical data from the ongoing EU Horizon 2020 CRADL (Continuous Regional Analysis Device for neonate Lung) project
registered at ClinicalTrials.gov (NCT02962505). 

\section{Homogenization theory based on Hashin-Shtrikman coated ellipsoids}
A brief review on the classical homogenization theory based on Hashin-Shtrikman coated ellipsoids is given in this section,
see \cite{Milton2002} for an in depth derivation of the corresponding results.
\subsection{Notation and conventions}
The following notation and conventions will be used below.
Classical electromagnetic theory is considered based on SI-units \cite{Jackson1999},
and with time convention $\eu^{\ju\omega t}$ for time harmonic fields where $\omega$ is the angular frequency. 
Let $\mu_0$, $\epsilon_0$ and ${\rm c}_0$ denote the permeability, the permittivity and
the speed of light in vacuum, respectively, and where ${\rm c}_0=1/\sqrt{\mu_0\epsilon_0}$.
A passive, homogeneous and isotropic dielectric material with complex valued relative permittivity $\epsilon=\epsilon^\prime-\ju\epsilon^{\prime\prime}$
and real valued conductivity $\sigma_\mrm{r}\geq 0$ has complex valued conductivity (or admittivity) $\sigma$ given by
$\sigma=\sigma_\mrm{R}+\ju\sigma_\mrm{I}$ where $\sigma_\mrm{I}=\omega\epsilon_0\epsilon^\prime$ represents the lossless capacitive part and 
$\sigma_\mrm{R}=\sigma_\mrm{r}+\omega\epsilon_0\epsilon^{\prime\prime}\geq 0$ includes both the conduction and the dielectric losses.
The complex valued conductivity is conveniently written as $\sigma=\sigma_\mrm{R}\left(1+\ju\eta\right)$ where $\eta=\cot\delta$ is the loss
cotangent (corresponding to the more commonly used loss tangent $\tan\delta=\sigma_\mrm{R}/\sigma_\mrm{I}=1/\eta$).
Both parameters $\sigma_\mrm{R}$ and $\sigma_\mrm{I}$ may depend on frequency, in which case they must also satisfy
the associated Kramers-Kronig relations, see \eg \cite{Nussenzveig1972,King2009}.
The cartesian unit vectors are denoted $(\hat{\bm{x}}_1,\hat{\bm{x}}_2,\hat{\bm{x}}_3)$.
Finally, the real and imaginary part and the complex conjugate of a complex number $\zeta$ are denoted 
$\Re\!\left\{\zeta\right\}$, $\Im\!\left\{\zeta\right\}$ and $\zeta^*$, respectively.

\subsection{The Hashin-Shtrikman coated ellipsoid assemblage}\label{sect:HSassemblage}


\begin{figure}[htb]
\begin{picture}(50,200)
\put(37,0){\makebox(150,190){\includegraphics[width=8cm]{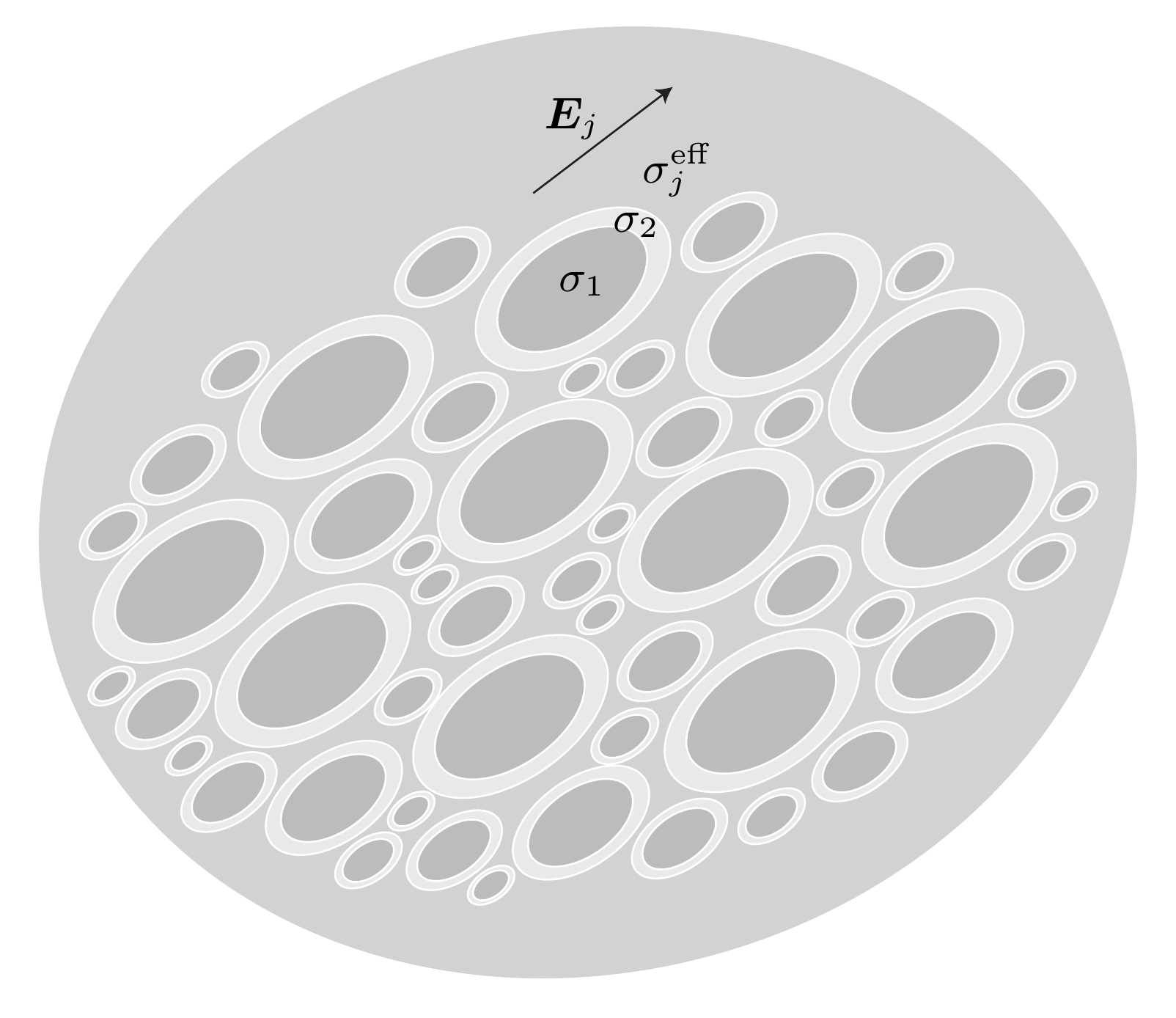}}} 
\end{picture}
\caption{The Hashin-Shtrikman coated ellipsoid assemblage. 
The figure illustrates a partial assemblage in a process that is completed when all space has been filled with ellipsoids.
Here, $\bm{E}_j$ denotes the applied electric field, $\sigma_j^\mrm{eff}$ the homogenized (effective) conductivity parameter and $\sigma_1$ and
$\sigma_2$ the core and the exterior conductivity of the prototype ellipsoid, respectively.}
\label{fig:HSellisoidsPDF}
\end{figure}

Consider a general anisotropic two-phase composite material consisting of an inclusion phase with conductivity $\sigma_1$ 
and an exterior phase with conductivity $\sigma_2$ and where the volume fraction of the inclusion phase is given by the parameter $f$.
A classical homogenization approach to model such a material is given by the Hashin-Shtrikman coated ellipsoid assemblage \cite{Milton2002}, 
as illustrated in figure \ref{fig:HSellisoidsPDF}.
All the coated ellipsoids are scaled and translated versions of a single prototype coated ellipsoid consisting
of a core (inclusion) phase with conductivity $\sigma_1$ and an exterior (coating) phase with conductivity $\sigma_2$.
The semi-axis lengths of the prototype core and exterior ellipsoids are denoted $l_{\mrm{c}_j}$ and
 $l_{\mrm{e}_j}$, respectively, and where $j=1,2,3$ refers to the cartesian coordinates $x_1$, $x_2$ and $x_3$, respectively.
 The prototype core and exterior ellipsoids are confocal in the sense that they can both be represented in the same system of elliptical coordinates as
 \begin{equation}\label{eq:ellipticalcoordinates}
 \frac{x_1^2}{c_1^2+\rho}+\frac{x_2^2}{c_2^2+\rho}+\frac{x_3^2}{c_3^2+\rho}=1,
 \end{equation}
 where $c_j$ are constants and $\rho$ the elliptical coordinate playing the role of the ``radius''.
 Hence, with $\rho_\mrm{c}$ and  $\rho_\mrm{e}$ denoting the ``radius'' of the two confocal 
 ellipsoids, we have
 \begin{equation}\label{eq:lcjlejdef}
 \frac{l_{\mrm{c}_j}^2}{c_j^2+\rho_\mrm{c}}=\frac{l_{\mrm{e}_j}^2}{c_j^2+\rho_\mrm{e}}=1,
 \end{equation}
 which implies that
 \begin{equation}\label{eq:lcjlejrel}
 l_{\mrm{e}_j}^2=l_{\mrm{c}_j}^2+\alpha, 
 \end{equation}
 where $\alpha=\rho_\mrm{e}-\rho_\mrm{c}>0$ and $j=1,2,3$.
 The volume of the two prototype ellipsoids are $V_\mrm{c}=4\pi l_{\mrm{c}_1}l_{\mrm{c}_2}l_{\mrm{c}_3}/3$ and
 $V_\mrm{e}=4\pi l_{\mrm{e}_1}l_{\mrm{e}_2}l_{\mrm{e}_3}/3$, and hence the volume fraction
 between the core phase and the total volume of the coated ellipsoid is given by
 \begin{equation}\label{eq:fdef}
 f=\frac{V_\mrm{c}}{V_\mrm{e}}=\frac{l_{\mrm{c}_1}l_{\mrm{c}_2}l_{\mrm{c}_3}}{l_{\mrm{e}_1}l_{\mrm{e}_2}l_{\mrm{e}_3}}.
 \end{equation}
 The Hashin-Shtrikman coated ellipsoid assemblage is obtained when the whole space is filled with 
scaled ellipsoids as indicated in figure \ref{fig:HSellisoidsPDF}.  
In this way, the resulting assemblage models the general anisotropic two-phase composite material where the volume fraction $f$ of the inclusion phase 
is preserved and parameterized according to \eref{eq:fdef}.
 
 \subsection{The effective conductivity}
 Consider a single prototype coated ellipsoid with core and exterior conductivities $\sigma_1$ and $\sigma_2$, respectively,
 as described in section \ref{sect:HSassemblage} above. Suppose further that the prototype ellipsoid is embedded in
 an arbitrary homogeneous auxiliary medium with conductivity $\sigma^\mrm{eff}$, and excited with an external static electric field
 $\bm{E}_j=E_0\hat{\bm{x}}_j$ aligned with the $j$th axis of the ellipsoid, as illustrated in figure \ref{fig:HSellisoidsPDF}. 
The fundamental equations to be solved are given by
 \begin{equation}\label{eq:fundamental}
\left\{\begin{array}{l}
\nabla\times\bm{E}(\bm{r})=\bm{0}, \vspace{0.2cm} \\
\nabla\cdot\bm{J}(\bm{r})=0, \vspace{0.2cm} \\
\bm{J}(\bm{r})=\sigma(\bm{r})\bm{E}(\bm{r}),
\end{array}\right.
\end{equation}
where $\bm{E}(\bm{r})$ and $\bm{J}(\bm{r})$ are the electric field intensity and the electric current density, respectively, and where 
$\sigma(\bm{r})$ is the complex valued conductivity which is assigned the appropriate constant values $(\sigma_1,\sigma_2,\sigma^\mrm{eff})$ 
inside and outside the prototype ellipsoid, respectively. 
The electric field outside the prototype ellipsoid may furthermore be denoted $\bm{E}(\bm{r})=\bm{E}_j(\bm{r})+\bm{E}_\mrm{s}(\bm{r})$
where $\bm{E}_\mrm{s}(\bm{r})$ may be thought of as the scattered field. 
The boundary conditions supplementing the equations in \eref{eq:fundamental} are obtained from
the continuity of the normal component of the current density $\bm{J}(\bm{r})$ at the media interfaces.
The equations in \eref{eq:fundamental} can be solved by introducing the scalar potential $\Phi(\bm{r})$ where $\bm{E}(\bm{r})=-\nabla\Phi(\bm{r})$, 
and where $\Phi(\bm{r})$ satisfies the Laplace equation $\nabla^2\Phi(\bm{r})=0$, together with the continuity of $\Phi(\bm{r})$
as well as the continuity of the normal current $\sigma(\bm{r})\frac{\partial}{\partial n}\Phi(\bm{r})$ at the media interfaces.

Since the elliptical coordinates constitute one of the very special coordinate systems for which the Laplace operator can be separated \cite{Morse+Feshbach1953b},
the electrostatic problem above can be solved analytically involving ordinary functions and elliptic integrals, see \eg \cite{Morse+Feshbach1953b,Bohren+Huffman1983,Milton2002}. 
In particular, by investigating these analytic solutions, it is found that one can choose the 
auxiliary medium parameter $\sigma^\mrm{eff}$ in such a way as to render the scattered field $\bm{E}_\mrm{s}(\bm{r})=\bm{0}$ 
(and at the same time there is a uniform non-zero field inside the ellipsoid core with the same polarization direction as the applied field).
Hence, in this situation the prototype ellipsoid is in some sense cloaked as it does not interfere with the surrounding uniform current field.
The resulting auxiliary, or effective, medium parameter $\sigma^\mrm{eff}_j$ depend in general on the polarization direction $\hat{\bm{x}}_j$, and is given by
\begin{equation}\label{eq:sigmaeffjdef}
\sigma^\mrm{eff}_j=\sigma_2+\frac{f\sigma_2\left(\sigma_1-\sigma_2\right)}{\sigma_2+\left(d_{\mrm{c}_j}-fd_{\mrm{e}_j}\right)\left(\sigma_1-\sigma_2\right)},
\end{equation}
where the depolarizing factors (\cf the geometrical factors in \cite{Bohren+Huffman1983}, and the demagnetizing factors in \cite{Osborn1945})  
$d_{\mrm{c}_j}$ and $d_{\mrm{e}_j}$ are given by $d_{\mrm{c}_j}=d_j(l_{\mrm{c}_1},l_{\mrm{c}_2},l_{\mrm{c}_3})$,
$d_{\mrm{e}_j}=d_j(l_{\mrm{e}_1},l_{\mrm{e}_2},l_{\mrm{e}_3})$ where
\begin{eqnarray}\label{eq:depolfactors}
d_j(l_1,l_2,l_3) \nonumber \\
\quad =\frac{l_1l_2l_3}{2}\int_{0}^{\infty}\frac{\mrm{d}y}{\left(l_j^2+y\right)\sqrt{\left(l_1^2+y\right)\left(l_2^2+y\right)\left(l_3^2+y\right)}},
\end{eqnarray}
and where $l_1$, $l_2$ and $l_3$ are the semi-axis lengths of the corresponding ellipsoids, and $j=1,2,3$, see \cite[p.~129]{Milton2002}.
The depolarizing factors are normalized in the sense that $d_1+d_2+d_3=1$.

Since further scaled and translated coated ellipsoids can be inserted into the effective medium without disturbing the surrounding uniform current field, 
the resulting Hashin-Shtrikman coated ellipsoid assemblage can finally be viewed from a macroscopic scale to have the 
homogeneous and anisotropic constitutive relation
\begin{equation}\label{eq:constJeqsigmaEdef}
\bm{J}=\bm{\sigma}^\mrm{eff}\cdot\bm{E},
\end{equation}
where the effective conductivity dyadic $\bm{\sigma}^\mrm{eff}$ is given by 
\begin{equation}\label{eq:sigmaeffdyaddef}
\bm{\sigma}^\mrm{eff}=\sum_{j=1}^{3}\sigma^\mrm{eff}_j\hat{\bm{x}}_j\hat{\bm{x}}_j.
\end{equation}

The coated sphere is a special case of the coated ellipsoid with $d_{\mrm{c}_j}=d_{\mrm{e}_j}=1/3$ for $j=1,2,3$, and hence
\begin{equation}\label{eq:sigmasphdef}
\sigma^\mrm{eff}_\mrm{sph}=\sigma_2+\frac{3f\sigma_2\left(\sigma_1-\sigma_2\right)}{3\sigma_2+\left(1-f\right)\left(\sigma_1-\sigma_2\right)},
\end{equation}
which is identical with the classical Maxwell-Garnett mixing formula, see \eg \cite{Sihvola1999,Milton2002}.

\subsection{Spheroidal inclusions}
Spheroids are particularly simple ellipsoidal shapes having rotational symmetry, and for which the corresponding
depolarizing factors as well as their surface areas can be expressed by explicit formulas involving simple functions.
Without loss of generality, it will be assumed here that the axis of rotation is defined by the $x_1$-axis.

A prolate spheroid is characterized by its semi-axis properties $l_1>l_2=l_3=l$, and hence with 
depolarizing factors $d_1<d_2=d_3=d$ where $d_1<1/3$ and $d=(1-d_1)/2$.
The eccentricity of the prolate spheroid is defined by
\begin{equation}\label{eq:varepsilonprolsphdef}
\varepsilon=\sqrt{1-\left(\frac{l}{l_1}\right)^2},
\end{equation}
and the corresponding depolarizing factor $d_1$ is given by
\begin{equation}\label{eq:d1prolsphdef}
d_1=\frac{1-\varepsilon^2}{\varepsilon^2}\left(\frac{1}{2\varepsilon}\ln\left(\frac{1+\varepsilon}{1-\varepsilon} \right)-1 \right),
\end{equation}
\cf \eg \cite[p.~132]{Milton2002} and \cite[pp.~352--354]{Osborn1945}. The surface area of the prolate spheroid is furthermore
given by
\begin{equation}\label{eq:Sprolsphdef}
S=2\pi l^2+2\pi\frac{ll_1}{\varepsilon}\arcsin\varepsilon,
\end{equation}
see \cite[p.~364]{Zwillinger2003}.

An oblate spheroid is characterized by its semi-axis properties $l_1<l_2=l_3=l$, and hence with 
depolarizing factors $d_1>d_2=d_3=d$ where $d_1>1/3$ and $d=(1-d_1)/2$.
The eccentricity of the oblate spheroid is defined by
\begin{equation}\label{eq:varepsilonoblsphdef}
\varepsilon=\sqrt{1-\left(\frac{l_1}{l}\right)^2},
\end{equation}
and the corresponding depolarizing factor $d_1$ is given by
\begin{equation}\label{eq:d1oblsphdef}
d_1=\frac{1}{\varepsilon^2}\left(1-\frac{\sqrt{1-\varepsilon^2}}{\varepsilon}\arcsin\varepsilon \right),
\end{equation}
\cf \eg \cite[ p.~132]{Milton2002} and \cite[pp.~352--354]{Osborn1945}. The surface area of the oblate spheroid is furthermore
given by
\begin{equation}\label{eq:Soblsphdef}
S=2\pi l^2+\pi\frac{l_1^2}{\varepsilon}\ln\left(\frac{1+\varepsilon}{1-\varepsilon} \right),
\end{equation}
see \cite[p.~364]{Zwillinger2003}.

\section{A parametric model for the changes in the conductivity of a lung during tidal breathing}\label{sect:parametricmodel}
A Hashin-Shtrikman homogenization approach based on \eref{eq:sigmaeffjdef} is considered to model the changes in the 
complex valued conductivity of a lung during tidal breathing.
The lung is modelled as a two-phase composite material where the air-filled alveoli constitute the inclusion phase with 
volume fraction $f$ and conductivity $\sigma_1=\ju\omega\epsilon_0$ corresponding to the electric displacement current in vacuum (similar as in air). 
The conductivity $\sigma_2$ of the exterior phase can be identified from some a priori information regarding 
the effective conductivity $\sigma_\mrm{EI}$ of an inflated lung. Hence, it is assumed that the conductivity of the inflated lung 
obtained from measurements such as in \cite{Gabriel+etal1996b} can be used as an effective conductivity of the lung
corresponding to a certain maximum volume fraction $f_\mrm{EI}$ at end-inspiration, and where the alveoli have a maximally extended spherical shape. 
The conductivity of the exterior phase can then be obtained by solving the Maxwell-Garnett equation \eref{eq:sigmasphdef}
with respect to $\sigma_2$ when the effective parameter $\sigma^\mrm{eff}_\mrm{sph}=\sigma_\mrm{EI}$  is given. 
This is equivalent to finding the roots of the following second order polynomial equation
\begin{eqnarray}\label{eq:MaxwellGarnettroots}
\sigma_2^2 2\left(1-f_\mrm{EI}\right)+\sigma_2\left(\sigma_1\left(1+2f_\mrm{EI}\right)-\sigma_\mrm{EI}\left(2+f_\mrm{EI}\right)\right) \nonumber \\
\quad \quad -\sigma_\mrm{EI}\left(1-f_\mrm{EI} \right)\sigma_1=0,
\end{eqnarray}
and where the root of physical interest has $\Re\{\sigma_2\}>0$ and $\Im\{\sigma_2\}>0$ 
(unless the exterior phase is a Drude material with $\Im\{\sigma_2\}<0$, etc.).

Tidal breathing is then considered with small changes in alveolar air-filling where the corresponding volume fraction $f$ changes from its maximum value
$f_\mrm{EI}$ at end-inspiration to its minimum value $f_\mrm{EE}$ at end-expiration.
Two fundamentally different physical modes of alveolar air-fillings are considered for the  tidal breathing.
\begin{itemize}
\item {\em Spherical shaped alveoli with fixed shape and varying surface area:} 
The alveoli are assumed to have a fixed spherical shape and the volume change is obtained
by a change of its radius. This implies that the surface area of the alveoli as well as the whole structure of the lung is stretched during the tidal breathing.
\item {\em Spheroidal shaped alveoli with varying shape and fixed surface area:}
The alveoli are assumed to have a varying spheroidal shape and the volume change is obtained
by a change of its spheroidal eccentricity while keeping its surface area fixed. In this mode, the volume
change of the alveoli is due solely to a change in the alveolar shape (prolongation or flattening of the spheroid) with a minor stretch in the lung structure.
\end{itemize}

With a tidal breathing based on spherical shaped alveoli (fixed shape and varying surface area) 
the change in conductivity is obtained from \eref{eq:sigmasphdef}
as $\Delta\sigma^\mrm{eff}_\mrm{sph}=\sigma^\mrm{eff}_\mrm{sph}(f)-\sigma^\mrm{eff}_\mrm{sph}(f_\mrm{EI})$
where $f$ denotes the volume fraction of air-filled alveoli during tidal breathing and $f_\mrm{EI}$ the corresponding value at end-inspiration.
Hence
\begin{eqnarray}\label{eq:Deltasigmasph}
\Delta\sigma^\mrm{eff}_\mrm{sph}=\frac{3f\sigma_2\left(\sigma_1-\sigma_2\right)}{3\sigma_2+\left(1-f\right)\left(\sigma_1-\sigma_2\right)} \nonumber \\
\quad -\frac{3f_\mrm{EI}\sigma_2\left(\sigma_1-\sigma_2\right)}{3\sigma_2+\left(1-f_\mrm{EI}\right)\left(\sigma_1-\sigma_2\right)},
\end{eqnarray}
where $f=f_\mrm{EI}-\Delta f$ and $\Delta f>0$. For a comparison with the spheroidal case below,
the Hashin-Shtrikman prototype core sphere is defined to have unit radius at maximal volume fraction $f_\mrm{EI}$
corresponding to the surface area $S_0=4\pi$.

With a tidal breathing based on spheroidal shaped alveoli (varying shape and fixed surface area) 
the change in conductivity is obtained from \eref{eq:sigmaeffjdef} and defined
by $\Delta\sigma_j^\mrm{eff}=\sigma_j^\mrm{eff}(f)-\sigma_j^\mrm{eff}(f_\mrm{EI})$.
The prototype core spheroids are furthermore assumed to be unit spheres at the maximal volume fraction $f_\mrm{EI}$ at end-inspiration, and
hence from  \eref{eq:sigmaeffjdef} and \eref{eq:sigmasphdef}
\begin{eqnarray}\label{eq:Deltasigma}
\Delta\sigma_j^\mrm{eff}=\frac{f\sigma_2\left(\sigma_1-\sigma_2\right)}{\sigma_2+\left(d_{\mrm{c}_j}-fd_{\mrm{e}_j}\right)\left(\sigma_1-\sigma_2\right)}\nonumber  \\
-\frac{3f_\mrm{EI}\sigma_2\left(\sigma_1-\sigma_2\right)}{3\sigma_2+\left(1-f_\mrm{EI}\right)\left(\sigma_1-\sigma_2\right)},
\end{eqnarray}
where $f=f_\mrm{EI}-\Delta f$ and $\Delta f>0$. Here, $d_{\mrm{c}_j}$ and $d_{\mrm{e}_j}$ are the depolarizing factors
of the Hashin-Shtrikman prototype core and exterior spheroids at volume fraction $f$, respectively.
Since the Hashin-Shtrikman prototype core spheroid coincides with the unit sphere at maximal volume fraction $f_\mrm{EI}$,
the following relation is obtained from \eref{eq:fdef}
\begin{equation}\label{eq:frel}
f_\mrm{EI}=\frac{1}{l_{\mrm{e}_1}l_{\mrm{e}_2}l_{\mrm{e}_3}},
\end{equation}
where the product $l_{\mrm{e}_1}l_{\mrm{e}_2}l_{\mrm{e}_3}$ is proportional to the volume of the prototype exterior spheroid.
Finally, the eccentricity $\varepsilon$ of the prototype core spheroid as defined in \eref{eq:varepsilonprolsphdef} or \eref{eq:varepsilonoblsphdef}
is used as a parameter to control the shape of the spheroid, as well as its volume fraction $f<f_\mrm{EI}$.

\subsection{The prototype core spheroids}\label{sect:parametricmodelcorespheroids}
Consider a prolate core spheroid with $l_{\mrm{c}_1}$ being the length of the semi-axis of rotation and
$l_\mrm{c}=l_{\mrm{c}_2}=l_{\mrm{c}_3}$ the length of the orthogonal axes.
Let the spheroidal eccentricity $\epsilon=\sqrt{1-t^2}$ be fixed, where $t=l_\mrm{c}/ l_{\mrm{c}_1}$, $0<t<1$ and $0<\varepsilon<1$.
The surface area of the prototype core spheroid is fixed at $S=4\pi$, and hence \eref{eq:Sprolsphdef} yields the equation
\begin{equation}\label{eq:Sprolsphdef2}
2\pi l_\mrm{c}^2+2\pi\frac{l_\mrm{c}l_{\mrm{c}_1}}{\varepsilon}\arcsin\varepsilon=4\pi,
\end{equation}
and which can be solved for $l_\mrm{c}$ to yield 
\begin{equation}
l_\mrm{c}=\sqrt{\frac{2\varepsilon t}{\varepsilon t+\arcsin\varepsilon}},
\end{equation}
and $l_{\mrm{c}_1}=l_\mrm{c}/t$. 
The corresponding depolarizing factor $d_{\mrm{c}_1}$ is given by \eref{eq:d1prolsphdef} and $d_\mrm{c}=(1-d_{\mrm{c}_1})/2$.

Consider similarly an oblate core spheroid with $l_{\mrm{c}_1}$ being the length of the semi-axis of rotation and
$l_\mrm{c}=l_{\mrm{c}_2}=l_{\mrm{c}_3}$ the length of the orthogonal axes.
Let the spheroidal eccentricity $\epsilon=\sqrt{1-t^2}$ be fixed, where $t= l_{\mrm{c}_1}/l_\mrm{c}$, $0<t<1$ and $0<\varepsilon<1$.
The surface area of the prototype core spheroid is fixed at $S=4\pi$, and hence \eref{eq:Soblsphdef} yields the equation
\begin{equation}\label{eq:Soblsphdef2}
2\pi l_\mrm{c}^2+\pi\frac{{l_{\mrm{c}_1}^2}}{\varepsilon}\ln\left(\frac{1+\varepsilon}{1-\varepsilon}\right)=4\pi,
\end{equation}
and which can be solved for $l_\mrm{c}$ to yield 
\begin{equation}
l_\mrm{c}=\sqrt{\frac{4\varepsilon}{2\varepsilon+t^2\ln\left(\frac{1+\varepsilon}{1-\varepsilon}\right)}},
\end{equation}
and $l_{\mrm{c}_1}=tl_\mrm{c}$. 
The corresponding depolarizing factor $d_{\mrm{c}_1}$ is given by \eref{eq:d1oblsphdef} and $d_\mrm{c}=(1-d_{\mrm{c}_1})/2$.

\subsection{The prototype external spheroids}\label{sect:parametricmodelexternalspheroids}
By increasing the eccentricity $\varepsilon>0$ of the prototype core spheroid while keeping its surface area fixed, its volume will decrease.
Hence, by defining the volume of the prototype exterior spheroid $V_\mrm{e}=4\pi l_{\mrm{e}_1}l_{\mrm{e}_2}l_{\mrm{e}_3}/3$ to be fixed, 
the corresponding volume fraction $f$ of the inclusion phase will decrease. 
The volume fraction $f$ for a given eccentricity $\varepsilon$ is hence obtained from \eref{eq:fdef} and \eref{eq:frel} as
\begin{equation}\label{eq:frel2}
 f=f_\mrm{EI}l_{\mrm{c}_1}l_\mrm{c}^2.
 \end{equation}
 
 To find the semi-axes of the prototype exterior spheroid at volume fraction $f$, 
 the  relations \eref{eq:lcjlejrel} of the confocal ellipsoids are now inserted into the following equation based on the definition \eref {eq:fdef}
 \begin{equation}
 f^2l_{\mrm{e}_1}^2l_{\mrm{e}_2}^2l_{\mrm{e}_3}^2=l_{\mrm{c}_1}^2l_{\mrm{c}_2}^2l_{\mrm{c}_3}^2,
 \end{equation}
 and which is equivalent to finding the real and positive root
of the algebraic equation
\begin{eqnarray}
\alpha^3+\alpha^2\left(l_{\mrm{c}_1}^2+2l_\mrm{c}^2 \right)+\alpha\left(2l_{\mrm{c}_1}^2l_\mrm{c}^2+l_\mrm{c}^4\right) \nonumber  \\
+l_{\mrm{c}_1}^2l_\mrm{c}^4\left(1-\frac{1}{f^2} \right)=0.
\end{eqnarray}
Once the correct real valued and positive root $\alpha$ has been identified, the semi-axes lengths $l_{\mrm{e}_j}$ are given by \eref{eq:lcjlejrel}.
The eccentricity and the depolarizing factors $d_{\mrm{e}_j}$ of the prototype exterior spheroid are now given
by \eref{eq:varepsilonprolsphdef} and \eref{eq:d1prolsphdef} for the prolate spheroid, or by
\eref{eq:varepsilonoblsphdef} and \eref{eq:d1oblsphdef} for the oblate.

\subsection{Sensitivity analysis for high contrast inclusions}\label{sect:sensitivityanalysis}
A first order Taylor series approximation of the conductivity changes in \eref{eq:Deltasigmasph} and \eref{eq:Deltasigma}  is given by
\begin{equation}\label{eq:DeltasigmaTaylor}
\Delta\sigma_j^\mrm{eff}\approx\left. \frac{\mrm{d} \sigma_j^\mrm{eff}}{\mrm{d}f} \right|_{f=f_\mrm{EI}}\mrm{d}f,
\end{equation}
where $\sigma_j^\mrm{eff}$ is given by \eref{eq:sigmaeffjdef}, and where the spherical case
\eref{eq:Deltasigmasph} is a special case of \eref{eq:Deltasigma} for which  $d_{\mrm{c}_j}=d_{\mrm{e}_j}=1/3$ and
\eref{eq:sigmasphdef} is used. Note that in the spheroidal case as defined above, the depolarizing factors $d_{\mrm{c}_j}$ and $d_{\mrm{e}_j}$
depend on $f$ via $\varepsilon$.

Assume that the conductivity $\sigma_1$ of the inclusion phase is very small and negligible in comparison to the conductivity $\sigma_2$ of the exterior phase.
The exact expression \eref{eq:sigmaeffjdef} can then be approximated by
\begin{equation}\label{eq:sigmaeffjdefapprox}
\sigma^\mrm{eff}_j \approx \sigma_2\left(\frac{1-d_{\mrm{c}_j}+f\left(d_{\mrm{e}_j}-1\right)}{1+fd_{\mrm{e}_j}-d_{\mrm{c}_j}} \right).
\end{equation}
Hence, both $\sigma^\mrm{eff}_j$ and its derivative
\begin{equation}\label{eq:sigmaeffjdefapproxder}
\frac{\mrm{d}\sigma_j^\mrm{eff}}{\mrm{d} f} \approx
\sigma_2\frac{\mrm{d}}{\mrm{d} f}
\left\{\frac{1-d_{\mrm{c}_j}+f\left(d_{\mrm{e}_j}-1\right)}{1+fd_{\mrm{e}_j}-d_{\mrm{c}_j}} \right\},
\end{equation}
are proportional to the complex valued conductivity $\sigma_2$ of the exterior medium, and where the
constant of proportionality is real valued. 
In particular, in the spherical case the following simple expressions are obtained
\begin{equation}\label{eq:sigmaeffjdefapproxsph}
\sigma^\mrm{eff}_\mrm{sph} \approx \sigma_2\left(\frac{2-2f}{2+f} \right),
\end{equation}
and
\begin{equation}\label{eq:sigmaeffjdefapproxdersph}
\frac{\mrm{d}\sigma_\mrm{sph}^\mrm{eff}}{\mrm{d} f}\approx
-\sigma_2\frac{6}{\left(2+f \right)^2}.
\end{equation}
Note that for the a priori effective conductivity $\sigma_\mrm{EI}$ of the inflated lung, \eref{eq:sigmaeffjdefapproxsph} yields
\begin{equation}\label{eq:sigmaeffjdefapproxsphEI}
\sigma_\mrm{EI} \approx \sigma_2\left(\frac{2-2f_\mrm{EI}}{2+f_\mrm{EI}} \right),
\end{equation}
which approximates the solution to \eref{eq:MaxwellGarnettroots}.

By writing $\sigma_2=\Re\{\sigma_2\}\left(1+\ju\eta \right)$, 
and by employing \eref{eq:DeltasigmaTaylor} and \eref{eq:sigmaeffjdefapproxder}, it is
concluded that
\begin{equation}\label{eq:Deltasigmabytheory}
\Delta\sigma_j^\mrm{eff}\approx\Re\{\Delta\sigma_j^\mrm{eff}\}\left(1+\ju\eta \right),
\end{equation}
where $\eta$ is the loss cotangent associated with the exterior phase having
complex valued conductivity $\sigma_2$. Note finally that \eref{eq:sigmaeffjdefapproxsphEI}
implies that
\begin{equation}\label{sigmaEIapproxResigmaEIpleta}
\sigma_\mrm{EI}\approx\Re\{\sigma_\mrm{EI}\}\left(1+\ju\eta \right),
\end{equation}
expressing that the loss cotangent of $\sigma_\mrm{EI}$ is approximately the same as of $\sigma_2$.

\section{Numerical examples and clinical data}

\subsection{Theoretical study of tidal breathing}\label{sect:numextheotidalbreathing}

A theoretical study on the changes in the complex valued conductivity of a lung during tidal breathing is described below.
As a prerequisite for this modeling an a priori estimate of the complex valued conductivity $\sigma_\mrm{EI}$ of an inflated lung is needed.
Here, the data is taken from \cite[Fig.~2e on p.~2257]{Gabriel+etal1996b} with $\sigma_\mrm{EI}=\sigma_\mrm{R}^\mrm{a}+\ju\omega \epsilon_0\epsilon_\mrm{r}^\mrm{a}$
where $\sigma_\mrm{R}^\mrm{a}=0.1$\unit{S/m} and $\epsilon_\mrm{r}^\mrm{a}=2000$ at $200$\unit{kHz} yielding
$\sigma_\mrm{EI}=0.1\left(1+\ju 0.22 \right)$.
The resulting conductivity of the exterior phase is obtained from \eref{eq:MaxwellGarnettroots} and is given by
$\sigma_2=0.6318\left(1+\ju 0.22 \right)$\unit{S/m}. 
The complex valued conductivity of the air-filled alveoli is given by $\sigma_1=\ju 1.1127\cdot 10^{-5}$\unit{S/m},
and which hence can be regarded negligible in comparison to the conductivity $\sigma_2$ of the exterior phase.

Another a priori input needed for this modeling is the range of volume fractions $[f_\mrm{EE},f_\mrm{EI}]$ during tidal breathing.
Consider \eg the lung volume of an adult male with a functional residual capacity of $2.3$\unit{l} and tidal volume $0.5$\unit{l}, \cf \cite{Ganong2003}.
Suppose further that the weight of the two lungs is about $0.8$\unit{kg} \cite{Molina+DiMaio2012}, 
corresponding approximately to $0.8$\unit{l} of blood and tissue. The following volume fractions are then obtained
\begin{equation}
\left\{\begin{array}{l}
f_\mrm{EE}=\displaystyle \frac{2.3}{2.3+0.8}=0.75, \vspace{0.2cm} \\
f_\mrm{EI}=\displaystyle \frac{2.8}{2.8+0.8}=0.78,
\end{array}\right.
\end{equation}
where the results have been rounded to two digits.

\begin{figure}[htb]
\begin{picture}(50,140)
\put(90,0){\makebox(50,130){\includegraphics[width=8cm]{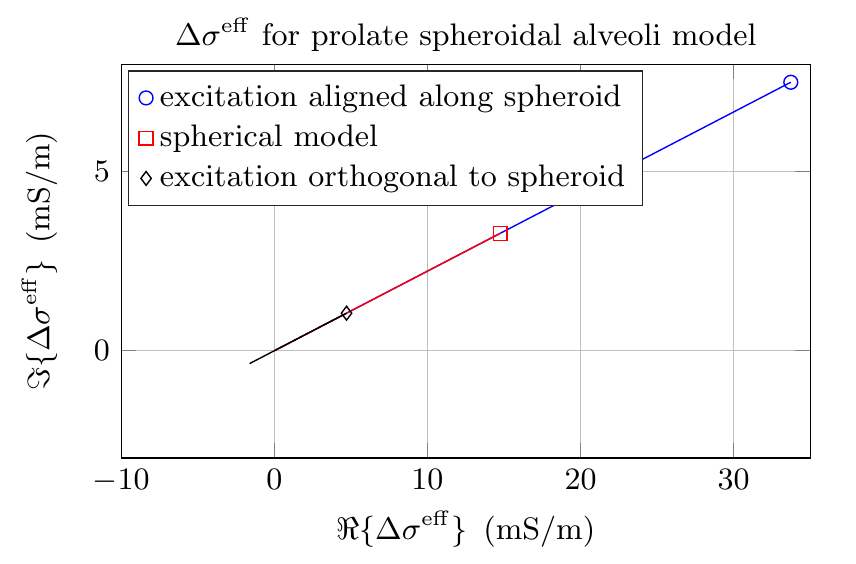}}} 
\end{picture}
\caption{Change in the effective lung conductivity parameter $\Delta\sigma^\mrm{eff}$,
plotted in the complex plane. The resulting change in complex conductivity $\Delta\sigma^\mrm{eff}=\sigma^\mrm{eff}(f)-\sigma^\mrm{eff}(f_\mrm{EI})$ is plotted for volume fractions ranging from
$f_\mrm{EI}=0.78$ to $f_\mrm{EE}=0.75$ where the last value at $f_\mrm{EE}$ is indicated with the corresponding plot symbol. The plot shows results with a prolate spheroidal alveoli (constant surface area and increasing eccentricity, hence decreasing volume) in comparison with a spherical alveoli (decreasing volume).}
\label{fig:matfig1}
\end{figure}
\begin{figure}[htb]
\begin{picture}(50,140)
\put(90,0){\makebox(50,130){\includegraphics[width=8cm]{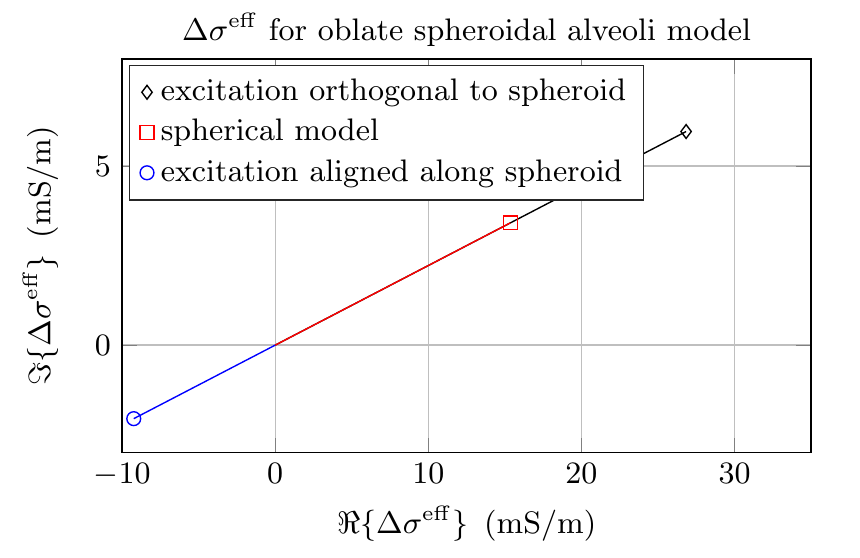}}} 
\end{picture}
\caption{Change in the effective lung conductivity parameter $\Delta\sigma^\mrm{eff}$,
plotted in the complex plane. The resulting change in complex conductivity $\Delta\sigma^\mrm{eff}=\sigma^\mrm{eff}(f)-\sigma^\mrm{eff}(f_\mrm{EI})$ is plotted for volume fractions ranging from
$f_\mrm{EI}=0.78$ to $f_\mrm{EE}=0.75$ where the last value at $f_\mrm{EE}$ is indicated with the corresponding plot symbol. The plot shows results with an oblate spheroidal alveoli (constant surface area and increasing eccentricity, hence decreasing volume) in comparison with a spherical alveoli (decreasing volume).}
\label{fig:matfig3}
\end{figure}
\begin{figure}[htb]
\begin{picture}(50,180)
\put(90,0){\makebox(50,170){\includegraphics[width=8cm]{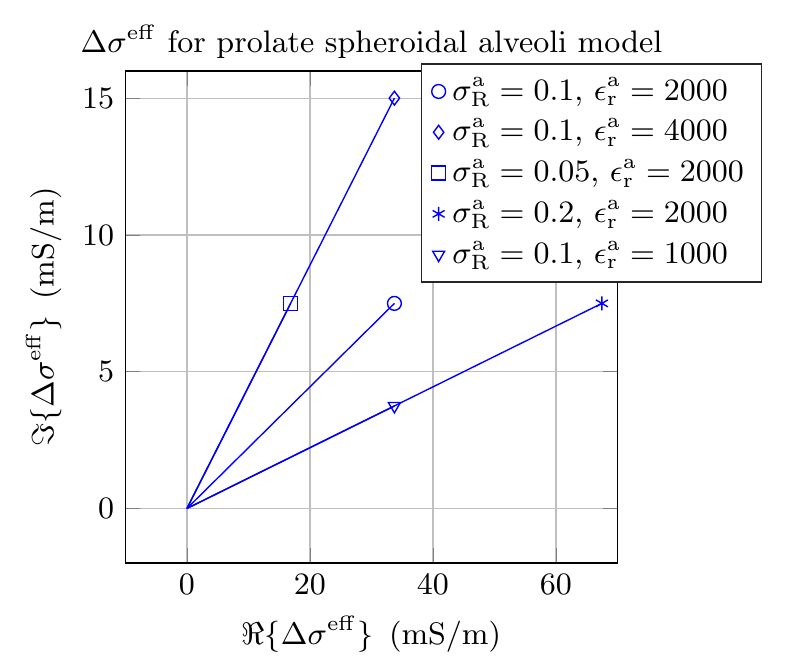}}} 
\end{picture}
\caption{Change in the effective lung conductivity parameter $\Delta\sigma^\mrm{eff}$ for prolate spheroidal alveoli 
with excitation aligned along the spheroid, as in figure \ref{fig:matfig1},
plotted here for different values of a priori inflated lung parameter $\sigma_\mrm{EI}=\sigma_\mrm{R}^\mrm{a}+\ju\omega \epsilon_0\epsilon_\mrm{r}^\mrm{a}$.}
\label{fig:matfig5}
\end{figure}

The theoretical study based on \eref{eq:Deltasigmasph} and \eref{eq:Deltasigma} (spheres and spheroids) 
is illustrated in figures \ref{fig:matfig1} through \ref{fig:matfig5} 
where the changes in conductivity are plotted in the complex plane and
parameterized by the volume fraction $f=f_\mrm{EI}-\Delta f$ with $\Delta f\in [0,0.03]$ and $f_\mrm{EI}=0.78$.
In figures \ref{fig:matfig1} and \ref{fig:matfig3}, it is noted that there is a fixed, almost linear relationship between the changes
in the real part of the conductivity $\Re\{\Delta\sigma^\mrm{eff}\}$ and the imaginary part
$\Im\{\Delta\sigma^\mrm{eff}\}$.
This observation is in full agreement with the theory predicted by the Taylor series approximation \eref{eq:DeltasigmaTaylor}
and \eref{eq:Deltasigmabytheory} assuming that the changes in volume fraction $\mrm{d}f=-\Delta f<0$ are small.
Hence, the linear slopes seen in figures \ref{fig:matfig1} and \ref{fig:matfig3} are consistent with the theory predicted by
\eref{eq:Deltasigmabytheory} and where $\eta= 0.22$ is the loss cotangent associated with the exterior phase having
complex valued conductivity $\sigma_2=0.6318\left(1+\ju 0.22 \right)$.
Note that this loss cotangent is also approximately the same as the one associated with the a priori (effective) background
conductivity $\sigma_\mrm{EI}=0.1\left(1+\ju 0.22 \right)$ according to the theory expressed in \eref{sigmaEIapproxResigmaEIpleta}.


As predicted by the linearization theory expressed in \eref{eq:sigmaeffjdefapproxder},
the linear slopes seen in figures \ref{fig:matfig1} and \ref{fig:matfig3} are independent of the assumed alveoli model 
(spherical, prolate spheroidal or oblate spheroidal),
as well as of the assumed excitation polarization (excitation aligned along spheroid or orthogonal to it).
It is only the magnitude of the conductivity changes that differ in between these different modes of alveolar air-filling and their anisotropy.


In figure \ref{fig:matfig5} is illustrated the different sensitivity slopes that are obtained 
with a prolate spheroidal alveoli model excitet along its symmetry axis, and with different a priori assumed background conductivities 
$\sigma_\mrm{EI}=\sigma_\mrm{R}^\mrm{a}+\ju\omega \epsilon_0\epsilon_\mrm{r}^\mrm{a}$.
As before, the resulting change in complex valued conductivity $\Delta\sigma^\mrm{eff}=\sigma^\mrm{eff}(f)-\sigma^\mrm{eff}(f_\mrm{EI})$ 
is plotted for volume fractions ranging from $f_\mrm{EI}=0.78$ to $f_\mrm{EE}=0.75$ where the last value at $f_\mrm{EE}$ 
is indicated with the corresponding plot symbol. 
Note that the corresponding changes $\Delta\sigma^\mrm{eff}$ are directly proportional to 
$\sigma_2$ by \eref{eq:sigmaeffjdefapproxder}, or $\sigma_\mrm{EI}$ by \eref{eq:sigmaeffjdefapproxsphEI}.

\subsection{Suboptimal imaging of permittivity based on EIT trained for real valued data}\label{sect:suboptimalEIT}
The clinical study that is referred to below has been performed within the EU-funded project CRADL, \cf ClinicalTrials.gov (NCT02962505).
The EIT voltage data from each neonatal patient is acquired using the CRADL device (Swisstom Landquart, Switzerland)
using a textile sensor belt around the chest comprising 32 electrodes, and measured at an operating frequency of about 200\unit{kHz}, down-converted and digitized at
a sampling rate of 48\unit{Hz}.
A four-point measurement strategy is employed using pairs of current stimulation and voltage measurement electrodes having an internal distance of 5 sensors.
To avoid amplifier saturation the measurements are omitted when measurement electrodes are at a distance smaller than or equal to 2 sensors from the stimulation electrodes.
The electronics of the measurement equipment allows for high precision in-phase and quadrature-phase (I/Q) measurements
and hence complex valued data is stored. 

%

The aim of the present study is to demonstrate that there is a non-negligible imaginary part of the data that may have a potential
for clinical use.  In particular, the theoretical study above show that
the change in the imaginary part of the conductivity is expected to be in the order of about 20\unit{\%} of the corresponding change in the real part
during tidal breathing ($\eta=0.22$ in the theoretical study above).

To start with, we will give a brief motivation to why the reconstruction matrix optimized using the GREIT algorithm 
and real valued data and parameters, is also expected to be (suboptimally) useful for reconstruction of the imaginary
part of the conductivity. Using the GREIT algorithm, a real valued reconstruction matrix $R$ is chosen to
optimize the criterion
\begin{eqnarray}\label{eq:realGREIT}
\begin{array}{llll}
	& \minimize & & \displaystyle \sum_k\| \Delta\sigma_\mrm{R}^{(k)}-R\Delta v^{(k)} \|_{W^{(k)}}^2  
\end{array}
\end{eqnarray}
where the index $k$ refers to a sequence of training data, 
$\Delta\sigma_\mrm{R}^{(k)}$ the chosen perturbations in the real valued conductivity,
$\Delta v^{(k)}$ the corresponding changes in the real valued voltage data obtained by using a forward model based on Finite Element Modeling (FEM)
and some particular stimulation patters, and $W^{(k)}$ a weighting matrix chosen according to some a priori weighting 
with respect to the training set, \cf \cite{Adler+etal2009}.
Assuming that the changes in conductivity are small, a first order Taylor series approximation at the background conductivity profile $\sigma_\mrm{R}$ 
can be used to express the changes as
\begin{eqnarray}\label{eq:GREITTaylorR}
\Delta v^{(k)}=v\left(\sigma_\mrm{R}+\Delta\sigma_\mrm{R}^{(k)}\right)-v\left(\sigma_\mrm{R}\right) \nonumber \\
\quad \approx \left.\frac{\partial v}{\partial \sigma}\right|_{\sigma_\mrm{R}}\Delta\sigma_\mrm{R}^{(k)},
\end{eqnarray}
where $\Delta v^{(k)}$ and $\Delta\sigma_\mrm{R}^{(k)}$ can be interpreted as column vectors, $v(\sigma_\mrm{R})$ is the forward model and 
$\left.\frac{\partial v}{\partial \sigma}\right|_{\sigma_\mrm{R}}$ the corresponding Jacobian matrix.

Suppose now that the background conductivity $\sigma_\mrm{R}$ is changed by adding an imaginary part so
that $\sigma=\sigma_\mrm{R}+\ju \sigma_\mrm{I}$. In a corresponding complex algorithm, 
a complex valued reconstruction matrix $R_\mrm{c}$ would now be chosen to optimize the criterion
\begin{eqnarray}\label{eq:complexGREIT}
\begin{array}{llll}
	& \minimize & & \displaystyle \sum_k\| \Delta\sigma^{(k)}-R_\mrm{c}\Delta v_\mrm{c}^{(k)} \|_{W^{(k)}}^2,
\end{array}
\end{eqnarray}
where $\Delta\sigma^{(k)}=\Delta\sigma_\mrm{R}^{(k)}+\ju \Delta\sigma_\mrm{I}^{(k)}$ denotes the 
chosen complex valued conductivity perturbations and $\Delta v_\mrm{c}^{(k)}$ the corresponding changes in the voltage data obtained by using the same forward model $v(\sigma)$.
Again, by assuming that the changes in conductivity are small in relation to the real valued background conductivity $\sigma_\mrm{R}$,
a first order Taylor series approximation at $\sigma_\mrm{R}$ can be used to express the changes as
\begin{eqnarray}\label{eq:GREITTaylorC}
\Delta v_\mrm{c}^{(k)}=v\left(\sigma_\mrm{R}+\ju\sigma_\mrm{I}+\Delta\sigma^{(k)}\right)-v\left(\sigma_\mrm{R}+\ju\sigma_\mrm{I}\right) \nonumber \\
\quad \approx  \left.\frac{\partial v}{\partial \sigma}\right|_{\sigma_\mrm{R}}\Delta\sigma^{(k)},
\end{eqnarray}
where the Jacobian matrix is the same as in \eref{eq:GREITTaylorR}.
Assuming that the same real valued perturbations $\Delta\sigma_\mrm{R}^{(k)}$ will be used as in \eref{eq:realGREIT}, 
we can write
\begin{equation}\label{eq:GREITDeltasigmak}
\Delta\sigma^{(k)}=\Delta\sigma_\mrm{R}^{(k)}\left(1+\ju\eta^{(k)} \right),
\end{equation}
where $\eta^{(k)}$ is a real valued parameter. Hence, by using \eref{eq:GREITTaylorR}, \eref{eq:GREITTaylorC} and \eref{eq:GREITDeltasigmak} it is seen that
\begin{eqnarray}
\Delta v_\mrm{c}^{(k)}\approx \left.\frac{\partial v}{\partial \sigma}\right|_{\sigma_\mrm{R}}\Delta\sigma_\mrm{R}^{(k)}\left(1+\ju\eta^{(k)} \right) \nonumber\\
\quad \approx\Delta v^{(k)}\left(1+\ju\eta^{(k)} \right),
\end{eqnarray}
which yields the following criterion based on \eref{eq:complexGREIT}
\begin{eqnarray}\label{eq:complexGREIT2}
\begin{array}{llll}
	& \minimize & & \displaystyle \sum_k \left| 1+\ju\eta^{(k)}\right|^2  \\
	&                 &  & \| \Delta\sigma_\mrm{R}^{(k)}-R_\mrm{c}\Delta v^{(k)} \|_{W^{(k)}}^2.
\end{array}
\end{eqnarray}

Now, according to the theoretical results of section \ref{sect:parametricmodel}, and in particular \eref{eq:Deltasigmabytheory} which are illustrated in section
\ref{sect:numextheotidalbreathing}, it is expected that the real and imaginary parts of the changes in conductivity of a lung during tidal breathing
will follow the linear relationship
\begin{equation}\label{eq:GREITDeltasigma}
\Delta\sigma=\Delta\sigma_\mrm{R}\left(1+\ju\eta \right),
\end{equation}
where $\eta$ is a real valued constant. 
This means that that the factor $1+\ju\eta^{(k)}$ in \eref{eq:complexGREIT2} should be independent of $k$ with $\eta^{(k)}=\eta$,
and hence that the optimal solution $R_\mrm{c}$ in \eref{eq:complexGREIT2} coincides with the optimal solution $R$ in \eref{eq:realGREIT}.
In conclusion, based on the assumptions of small changes with linear approximations as in \eref{eq:GREITTaylorR} and \eref{eq:GREITTaylorC},
together with training sequences $\Delta\sigma^{(k)}$ following the linear relationship \eref{eq:GREITDeltasigma}, 
it is expected that the choice $R_\mrm{c}=R$ will yield a suboptimal solution to \eref{eq:complexGREIT} that can be useful also for the
reconstruction of the imaginary part of the complex valued conductivity.


\subsection{A comparison with clinical data}\label{sect:CRADLexamples}



\begin{figure}[htb]
\begin{picture}(50,110)
\put(90,0){\makebox(50,100){\includegraphics[width=8cm]{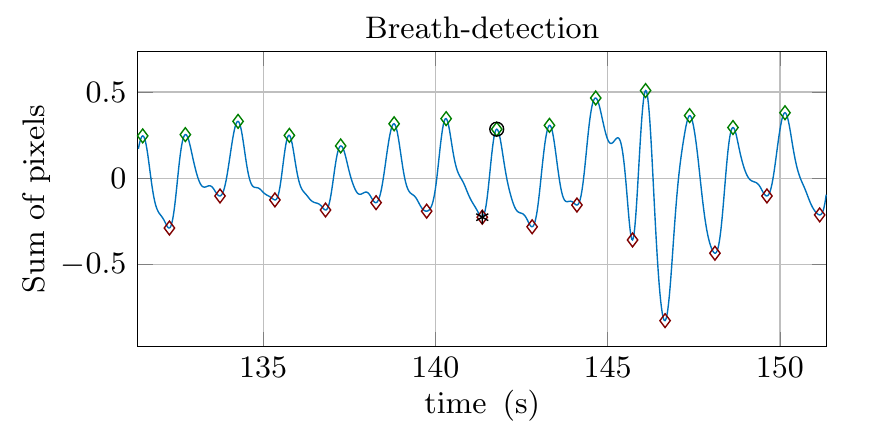}}} 
\end{picture}
\caption{Breath detection based on the breathing signal (sum of image pixel values). The lower 
red and upper green diamonds indicate the timing of end-expiration and end-inspiration, respectively. 
The black star and circle in the middle of the figure indicate breath no 88 corresponding to the tidal images shown in figure \ref{fig:CP1br88}.}
\label{fig:CP1matfig1br88}
\end{figure}
\begin{figure}[t]
\begin{picture}(50,200)
\put(90,0){\makebox(50,190){\includegraphics[width=8cm]{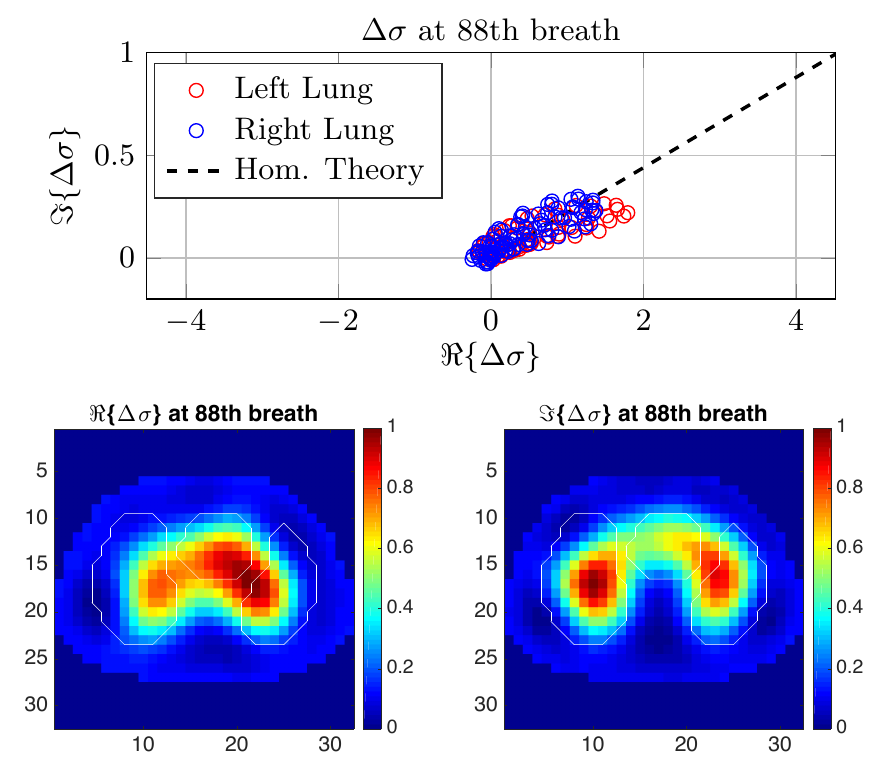}}} 
\end{picture}
\caption{Reconstructed pixel values of tidal image at breath no.~88. The upper plot shows a comparison with the presented homogenization theory (dashed line),
and the lower plots show normalized tidal images of the real and imaginary conductivity changes, respectively, in arbitrary units.}
\label{fig:CP1br88}
\end{figure}
\begin{figure}[t]
\begin{picture}(50,200)
\put(90,0){\makebox(50,190){\includegraphics[width=8cm]{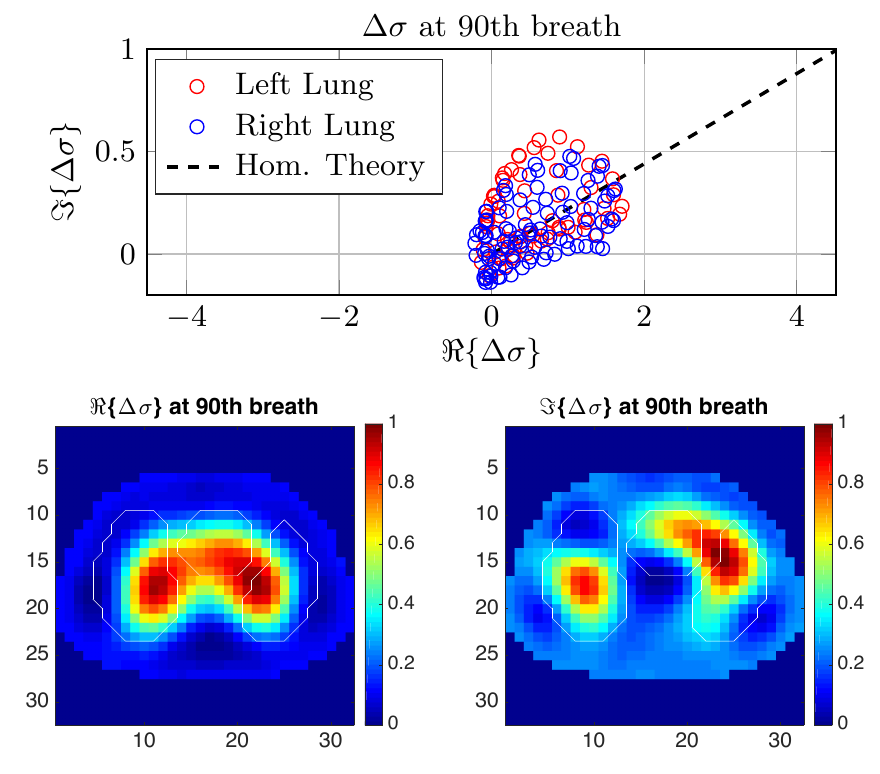}}} 
\end{picture}
\caption{Same as in figure \ref{fig:CP1br88}, at breath no.~90.}
\label{fig:CP1br90}
\end{figure}
\begin{figure}[t]
\begin{picture}(50,200)
\put(90,0){\makebox(50,190){\includegraphics[width=8cm]{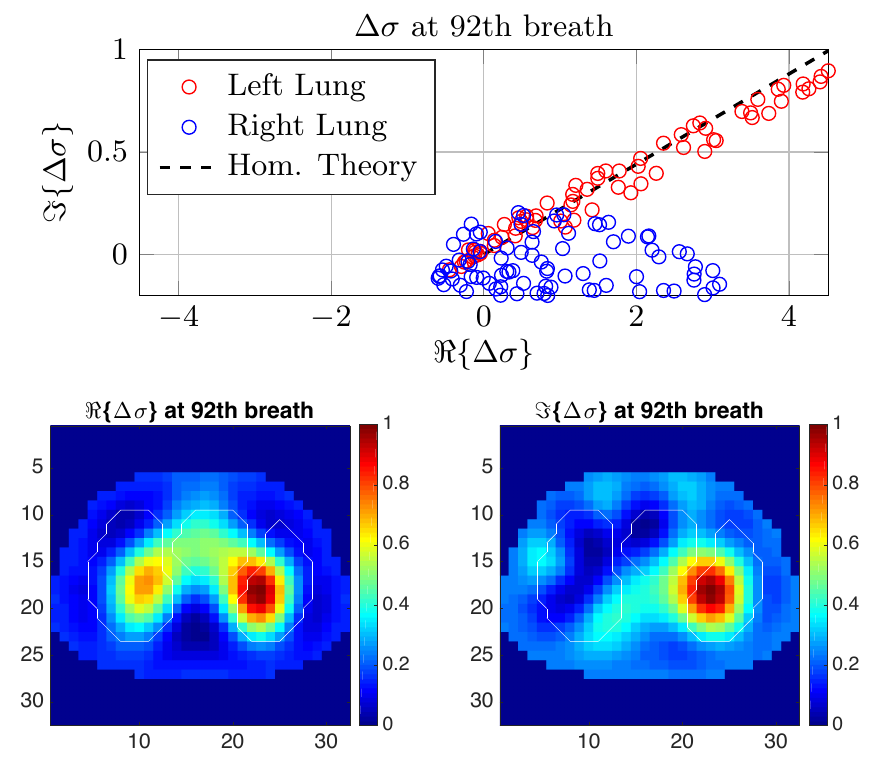}}} 
\end{picture}
\caption{Same as in figure \ref{fig:CP1br88}, at breath no.~92.}
\label{fig:CP1br92}
\end{figure}
\begin{figure}[htb]
\begin{picture}(50,110)
\put(90,0){\makebox(50,100){\includegraphics[width=8cm]{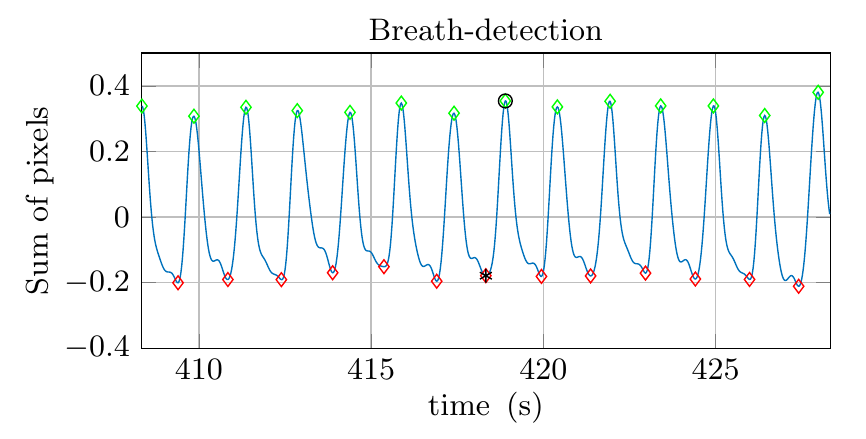}}} 
\end{picture}
\caption{Breath detection based on the breathing signal (sum of image pixel values). The lower 
red and upper green diamonds indicate the timing of end-expiration and end-inspiration, respectively. 
The black star and circle in the middle of the figure indicate breath no 262 corresponding to the tidal images shown in figure \ref{fig:CP1br262}.}
\label{fig:CP1matfig1br262}
\end{figure}
\begin{figure}[t]
\begin{picture}(50,200)
\put(90,0){\makebox(50,190){\includegraphics[width=8cm]{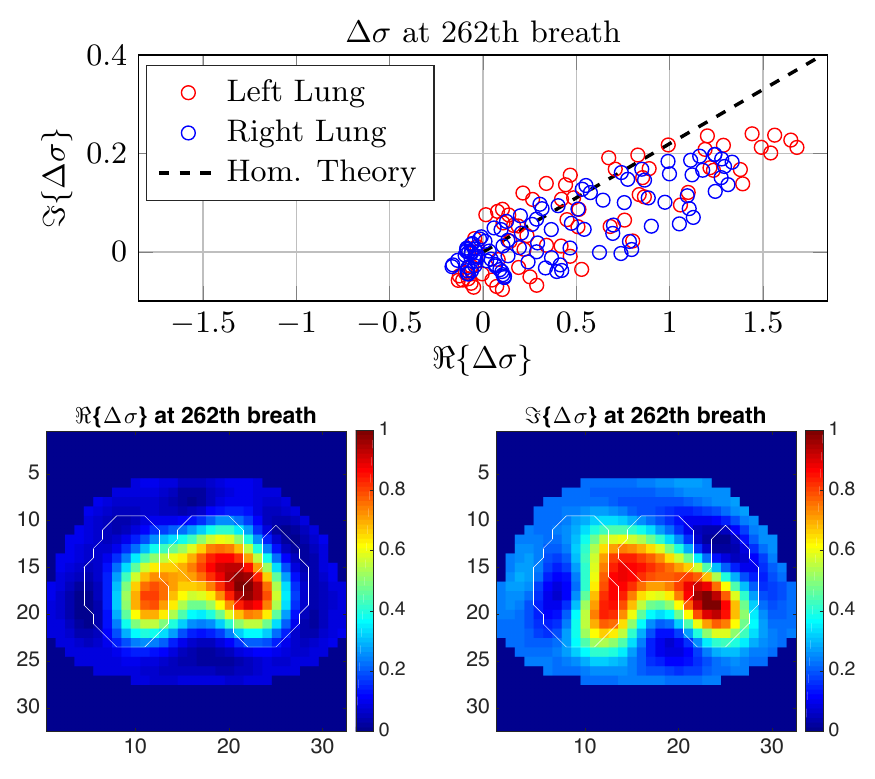}}} 
\end{picture}
\caption{Reconstructed pixel values of tidal image at breath no.~262. The upper plot shows a comparison with the presented homogenization theory (dashed line),
and the lower plots show normalized tidal images of the real and imaginary conductivity changes, respectively, in arbitrary units.}
\label{fig:CP1br262}
\end{figure}
\begin{figure}[t]
\begin{picture}(50,200)
\put(90,0){\makebox(50,190){\includegraphics[width=8cm]{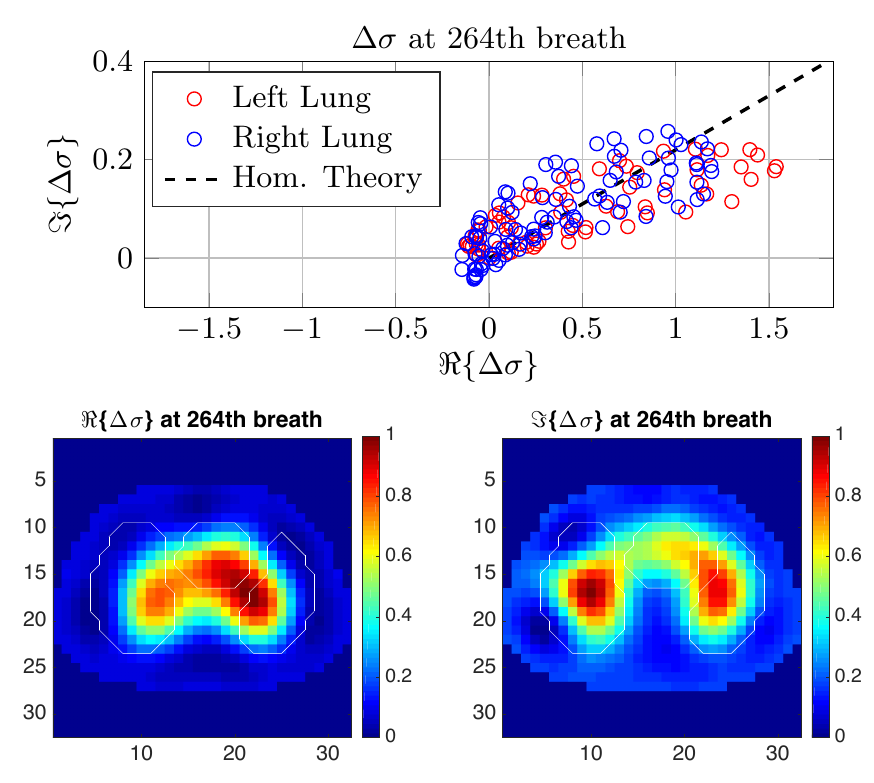}}} 
\end{picture}
\caption{Same as in figure \ref{fig:CP1br262}, at breath no.~264.}
\label{fig:CP1br264}
\end{figure}
\begin{figure}[t]
\begin{picture}(50,200)
\put(90,0){\makebox(50,190){\includegraphics[width=8cm]{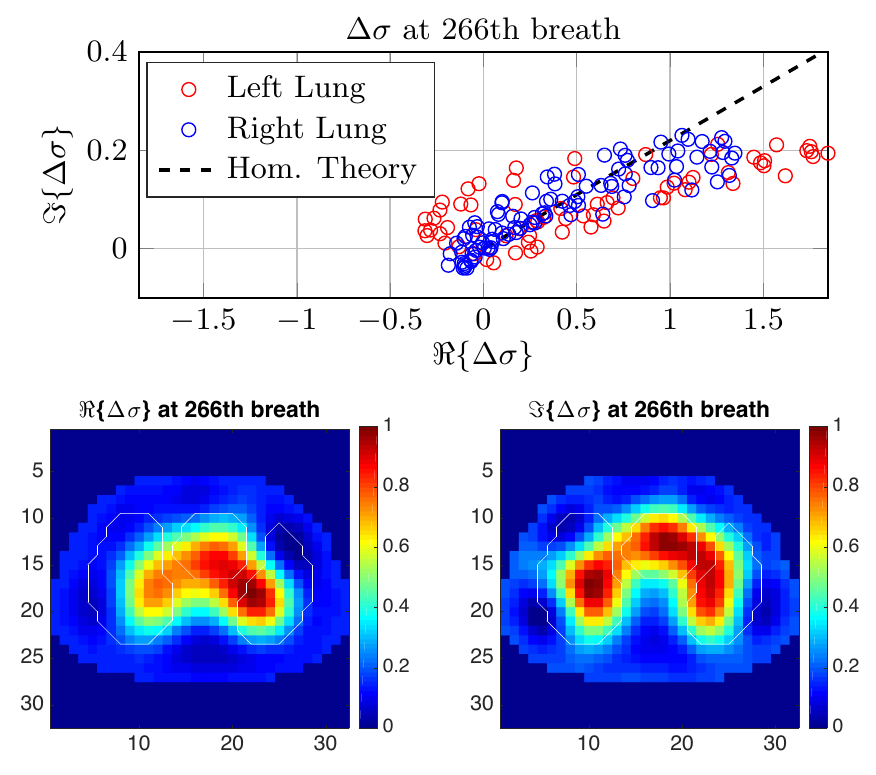}}} 
\end{picture}
\caption{Same as in figure \ref{fig:CP1br262}, at breath no.~266.}
\label{fig:CP1br266}
\end{figure}

As an illustration of its potential clinical use, the presented theory is employed here in a comparison with a small set of clinical
data collected within the CRADL project, as mentioned above.
The EIT data from one mechanically ventilated neonatal patient (weight 1400\unit{g} and gestational age 29 weeks)
is included in the examples below, as shown in Figures \ref{fig:CP1matfig1br88} through \ref{fig:CP1br266}.
Two sequences of data have been included in these examples,
one with a somewhat unstable breathing period shown in figures \ref{fig:CP1matfig1br88} through \ref{fig:CP1br92},
and one with a more stable ventilation as shown in figures \ref{fig:CP1matfig1br262} through \ref{fig:CP1br266}.
The figures \ref{fig:CP1matfig1br88} and \ref{fig:CP1matfig1br262} illustrate the corresponding breathing signals comprising the sum of all reconstructed image pixels 
of real valued conductivity changes, and which is the input for the breath detection \cite{Khodadad+etal2017}. 

Based on the identified timing of end-expiration 
and end-inspiration, a tidal image at a particular breath is then obtained by applying a precalculated real valued
reconstruction matrix to the corresponding voltage difference data (end-expiration data minus end-inspiration data), as described in section \ref{sect:suboptimalEIT}.
Figures \ref{fig:CP1br88} through \ref{fig:CP1br92}, and \ref{fig:CP1br262} through \ref{fig:CP1br266}
show the reconstructed pixel values and tidal images at breath no.~88-92 and 262-266, respectively, 
for both the real and the imaginary part of the complex valued conductivity changes.
The upper figures show the complex valued pixels in a comparison with the presented homogenization theory of section \ref{sect:parametricmodel},
and in particular \eref{eq:Deltasigmabytheory} where $\Delta\sigma=\Re\{\Delta\sigma\}\left(1+\ju\eta \right)$ and where $\eta=0.22$ (dashed line), 
as described in section \ref{sect:numextheotidalbreathing}. 
The red and blue circles indicate  pixels from the left lung and the right lung, respectively. 
The lower plots show tidal images of the real and the imaginary parts of the reconstructed conductivity changes, respectively, 
with the corresponding contours of the left lung, the right lung and the heart indicated within the figure (the right lung contour appears
on the left-hand side of the figures, etc.).

As can be seen from these figures, the reconstructed pixel values agree quite well with the theory that predicts the linear relationship
$\Delta\sigma=\Re\{\Delta\sigma\}\left(1+\ju\eta \right)$ for a lung during tidal breathing.
On the other hand, we will also stress here that these images have been chosen deliberately with the purpose to illustrate the theory, 
and many other tidal images can not be interpreted in the same favorable way. 
However, one should also keep in mind
that there can be external disturbances and/or other competing conductivity mechanisms that are present
during the measurements. 

The figure \ref{fig:CP1br92} (at breath no.~92) is particularly interesting, as it is
only the left lung pixels that show a clear linear ``correlation'' as predicted by theory, and it is also only the left lung
that appears in full contrast in the imaging of both the real and the imaginary part of the conductivity change. 
A hypothetical interpretation of this image is that it is only the left lung that is properly ventilated during this particular breath, 
and what is seen in the real valued conductivity change in the right lung could instead be caused by a change in blood perfusion, gas concentration, 
muscle contractions, etc., without much alveoli ventilation.

\subsection{Possible clinical applications}
The main conclusion from the theoretical analysis in this paper is that
the loss cotangent of the change in the complex valued conductivity of a lung during tidal breathing
is expected to be the same as of the effective background conductivity.
Hence, the presence of a linear relationship between the corresponding real and imaginary components could 
serve as part of a ``quality indicator'' for an EIT image, \ie to determine if the image can be trusted as representing 
primarily tidal ventilation and not some sort of artifact.
As was indicated above with reference to figure \ref{fig:CP1br92}, 
this observation could also be used as an indicator to distinguish when the right and/or the left lung is properly ventilated in contrast to
other competing physiological mechanisms of conductivity change (blood perfusion, muscle contractions, etc) and which are not expected 
to follow the same linear law as with the normal tidal breathing.

Another potentially interesting application is based on the assumption that the actual background conductivity of the lung is correlated 
with pulmonary fluid content.
Pulmonary edema, which is a clinically relevant pathology both in adults and neonates, is characterized by an increase in pulmonary fluid content \cite{Trepte+etal2016}. 
Hence, a change in the loss cotangent of the linear relationship between the real and imaginary part might be clinically useful for the detection of pulmonary edema.

\subsection{Uncertainties}
One parameter of major uncertainty in this study is the actual volume fraction $f$ of air during tidal breathing.
However, since there is a very high contrast between the complex valued conductivity of air and tissue,
the sensitivity analysis of section \ref{sect:sensitivityanalysis} is valid
showing that the result \eref{eq:Deltasigmabytheory} is independent of $f$.
Furthermore, the loss cotangent $\eta$ of $\sigma_2=\Re\{\sigma_2\}\left(1+\ju\eta \right)$
is essentially the same as the loss cotangent of the a priori parameter $\sigma_\mrm{EI}$, 
as expressed in \eref{eq:sigmaeffjdefapproxsphEI} and
\eref{sigmaEIapproxResigmaEIpleta}.
Hence, an uncertainty in the volume fraction $f_\mrm{EI}$ at end-inspiration (where $\sigma_\mrm{EI}$ is given) will only 
affect the magnitude of the estimated parameter $\sigma_2$,
and it will not affect the estimated loss cotangent $\eta$ nor the conclusion made in relation to \eref{eq:Deltasigmabytheory}.

Another uncertainty related to the measured CRADL EIT data described in section \ref{sect:CRADLexamples} above
is due to the fact that the measured I and Q data have been rotated already in the hardware in order to minimize the imaginary (Q) part of the data.
The main purpose of this procedure is to minimize the effect of the phase shifts in the internal electronics as well as in the stray capacitances in leads and electrodes, etc.
However, this means also that there is an unknown rotation angle $\phi$ present in the data relating to a ``true'' resistive background. 
Notably, since the reconstruction matrix in our case is real valued,
the presence of the unknown rotation angle $\phi$ means simply that our actual estimates are also rotated as
\begin{equation}
\left\{\begin{array}{l}
\sigma_\mrm{R}=\sigma_\mrm{R}^\mrm{t}\cos\phi-\sigma_\mrm{I}^\mrm{t}\sin\phi, \vspace{0.2cm} \\
\sigma_\mrm{I}=\sigma_\mrm{R}^\mrm{t}\sin\phi+\sigma_\mrm{I}^\mrm{t}\cos\phi,
\end{array}\right.
\end{equation}
where $\sigma_\mrm{R}^\mrm{t}$ and $\sigma_\mrm{I}^\mrm{t}$ are the desired ``true'' estimates that would have resulted 
if the calibration would have been carried out by using a ``true'' resistive phantom.
Obviously, the two sets of estimates $\{\sigma_\mrm{R},\sigma_\mrm{I}\}$ and $\{\sigma_\mrm{R}^\mrm{t},\sigma_\mrm{I}^\mrm{t}\}$ are equivalent 
in the sense of the linear and bijective rotation operation. However, for an EIT instrumentation that is designed to image the real valued conductivity, it is desirable to have $\phi$ as small as possible.
Hence, in a future investigation to exploit the complex valued data as described above,  a thorough
calibration procedure should be included to eliminate the phase uncertainty in the complex valued measurements.

\section{Summary and conclusions}
A theoretical model based on classical homogenization theory and the Hashin-Shtrikman coated ellipsoids has been derived to model the changes in the complex valued conductivity 
of a lung during tidal breathing. Here, the alveolar air-filling corresponds to the inclusion phase of the two-phase composite material. 
The model predicts a linear relationship between the real and the imaginary parts of the changes in the complex valued conductivity of a lung during tidal breathing,
and where the loss cotangent of the change is approximately the same as of the effective background conductivity.
Hence, even though the magnitude of the change depend on the chosen ellipsoidal model, 
the loss cotangent of the change is virtually independent of the shape and orientation (anisotropy) of the alveoli.
The theory has been illustrated with numerical examples, as well as by using reconstructed EIT images 
based on measurement data from the ongoing CRADL study. The new theory may have a potential for a development of new improved image reconstruction
algorithms exploiting complex valued measurement data, and/or to define new clinically useful outcome parameters in lung EIT.


\ack

This project has received funding from the European UnionÕs Horizon 2020 research and innovation programme under grant agreement No. 668259 (CRADL project).
The work has also been supported by the Swedish Foundation for Strategic Research (SSF) under the programme Applied Mathematics and the project Complex Analysis and Convex Optimization for EM Design.

\section*{References}


\begin{thebibliography}{10}

\bibitem{Frerichs2000}
I.~Frerichs.
\newblock Electrical impedance tomography {(EIT)} in applications related to
  lung and ventilation: a review of experimental and clinical activities.
\newblock {\em Physiological Measurement}, 15(2):1, 2000.

\bibitem{Carlisle2010}
H.~R. Carlisle, R.~K. Armstrong, P.~G. Davis, A.~Schibler, I.~Frerichs, and
  D.~G. Tingay.
\newblock Regional distribution of blood volume within the preterm infant
  thorax during synchronised mechanical ventilation.
\newblock {\em Intensive Care Med}, 36:2101--2108, 2010.

\bibitem{Frerichs+etal2017}
I.~Frerichs, M.~B.~P. Amato, A.~H. van Kaam, D.~G. Tingay, Z.~Zhao,
  B.~Grychtol, M.~Bodenstein, H.~Gagnon, S.~H. B\"{o}hm, E.~Teschner,
  O.~Stenqvist, T.~Mauri, V.~Torsani, L.~Camporota, A.~Schibler, G.~K. Wolf,
  D.~Gommers, S.~Leonhardt, and A.~Adler.
\newblock Chest electrical impedance tomography examination, data analysis,
  terminology, clinical use and recommendations: consensus statement of the
  {TR}anslational {EIT} developme{N}t stu{D}y group.
\newblock {\em Thorax}, 72:83--93, 2017.
\newblock doi:10.1136/thoraxjnl-2016-208357.

\bibitem{Somersalo+Cheney+Isaacson1992}
Erkki Somersalo, Margaret Cheney, and David Isaacson.
\newblock Existence and uniqueness for electrode models for electric current
  computed tomography.
\newblock {\em SIAM J. Appl. Math.}, 52(4):1023--1040, 1992.

\bibitem{Cheney+etal1999}
M.~Cheney, D.~Isaacson, and J.~C. Newell.
\newblock Electrical impedance tomography.
\newblock {\em SIAM Review}, 41(1):85--101, 1999.

\bibitem{Bayford2006}
R.~Bayford.
\newblock Bioimpedance tomography (electrical impedance tomography).
\newblock {\em Annu. Rev. Biomed. Eng.}, 8:63--91, 2006.

\bibitem{Nordebo+etal2010b}
S.~Nordebo, R.~Bayford, B.~Bengtsson, A.~Fhager, M.~Gustafsson, P.~Hashemzadeh,
  B.~Nilsson, T.~Rylander, and T.~Sj\"{o}d\'{e}n.
\newblock An adjoint field approach to {F}isher information-based sensitivity
  analysis in electrical impedance tomography.
\newblock {\em Inverse Problems}, 26, 2010.
\newblock 125008.

\bibitem{Adler+Guardo1996}
A.~Adler and R.~Guardo.
\newblock Electrical impedance tomography: regularized imaging and contrast
  detection.
\newblock {\em IEEE Transactions on Medical Imaging}, 15(2):170--179, 1996.

\bibitem{Adler+etal2009}
A.~Adler, J.~H.~Arnold abd R.~Bayford, A.~Borsic, B.~Brown, P.~Dixon, T.~J.~C.
  Faes, I.~Frerichs, H.~Gagnon, Y.~G\"{a}rber, B.~Grychtol, G.~Hahn, W.~R.~B.
  Lionheart, A.~Malik, R.~P. Patterson, J.~Stocks, A.~Tizzard, N.~Weiler, and
  G.~K. Wolf.
\newblock {GREIT}: a unified approach to {2D} linear {EIT} reconstruction of
  lung images.
\newblock {\em Physiological Measurement}, 30(6):35--55, 2009.

\bibitem{Khodadad+etal2017}
D.~Khodadad, S.~Nordebo, N.~Seifnaraghi, A.~D. Waldmann, B.~M\"{u}ller, and
  R.~Bayford.
\newblock Breath detection using short-time {F}ourier transform analysis in
  {E}lectrical {I}mpedance {T}omography.
\newblock In {\em 32nd URSI General Assembly \& Scientific Symposium}, pages
  2185--2187, Montreal, August 2017.

\bibitem{Cheney+Isaacson1995a}
Margaret Cheney and David Isaacson.
\newblock Issues in electrical impedance imaging.
\newblock {\em IEEE Computational Science \& Engineering}, pages 53--62, 1995.

\bibitem{Adler+etal1997}
A.~Adler, R.~Amyot, R.~Guardo, J.~H.~T. Bates, and Y.~Berthiaume.
\newblock Monitoring changes in lung air and liquid volumes with electrical
  impedance tomography.
\newblock {\em Journal of Applied Physiology}, 83(5):1762--1767, 1997.

\bibitem{Gabriel+etal1996b}
S.~Gabriel, R.~W. Lau, and C.~Gabriel.
\newblock The dielectric properties of biological tissues: {II}. {M}easurements
  in the frequency range 10 {Hz} to 20 {GHz}.
\newblock {\em Phys. Med. Biol.}, 41:2251--2269, 1996.

\bibitem{Isaacson+etal2006}
D.~Isaacson, J.~L. Mueller, J.~C. Newell, and S.~Siltanen.
\newblock Imaging cardiac activity by the {D}-bar method for electrical
  impedance tomography.
\newblock {\em Physiol. Meas.}, 27:43--50, 2006.

\bibitem{Hamilton+etal2017}
S.~J. Hamilton, J.~L. Mueller, and M.~Alsaker.
\newblock Incorporating a spatial prior into nonlinear {D}-bar {EIT} imaging
  for complex admittivities.
\newblock {\em IEEE Transactions on Medical Imaging}, 36(2):457--466, 2017.

\bibitem{Gabriel+etal1996a}
C.~Gabriel, S.~Gabriel, and E.~Corthout.
\newblock The dielectric properties of biological tissues: {I}. {L}iterature
  survey.
\newblock {\em Phys. Med. Biol.}, 41:2231--2249, 1996.

\bibitem{Gabriel+etal1996c}
S.~Gabriel, R.~W. Lau, and C.~Gabriel.
\newblock The dielectric properties of biological tissues: {III}. {P}arametric
  models for the dielectric spectrum of tissues.
\newblock {\em Phys. Med. Biol.}, 41:2271--2293, 1996.

\bibitem{Ganong2003}
William~F. Ganong.
\newblock {\em Review of medical physiology}.
\newblock Lange Medical Books/McGraw-Hill, New York, 21st edition, 2003.

\bibitem{Molina+DiMaio2012}
D.~K. Molina and V.~J. DiMaio.
\newblock Normal organ weights in men: part {II}-the brain, lungs, liver,
  spleen, and kidneys.
\newblock {\em Am J Forensic Med Pathol.}, 33:368--372, 2012.

\bibitem{Nopp+etal1993}
P.~Nopp, E.~Rapp, H.~Pf\"{u}tzner, H.~Nakesch, and C.~Ruhsam.
\newblock Dielectric properties of lung tissue as a function of air content.
\newblock {\em Phys. Med. Biol.}, 38:699--716, 1993.

\bibitem{Nopp+etal1997}
P.~Nopp, N.~D. Harris, T.~X. Zhao, and B.~H. Brown.
\newblock Model for the dielectric properties of human lung tissue against
  frequency and air content.
\newblock {\em Medical \& Biological Engineering \& Computing}, 35(6):695--702,
  1997.

\bibitem{Wang2014}
J-R Wang, B-Y Sun, H-X Wang, S~Pang, X~Xu, and Q~Sun.
\newblock Experimental study of dielectric properties of human lung tissue in
  vitro.
\newblock {\em Journal of Medical and Biological Engineering}, 34(6):598--604,
  2014.

\bibitem{Sihvola1999}
Ari Sihvola.
\newblock {\em Electromagnetic Mixing Formulae and Applications}.
\newblock {IEE} Electromagnetic Waves Series, 47. Institution of Electrical
  Engineers, 1999.

\bibitem{Milton2002}
Graeme~W. Milton.
\newblock {\em The Theory of Composites}.
\newblock Cambridge University Press, Cambridge, U.K., 2002.

\bibitem{Jackson1999}
J.~D. Jackson.
\newblock {\em Classical Electrodynamics}.
\newblock John Wiley \& Sons, New York, third edition, 1999.

\bibitem{Nussenzveig1972}
H.~M. Nussenzveig.
\newblock {\em Causality and dispersion relations}.
\newblock Academic Press, London, 1972.

\bibitem{King2009}
F.~W. King.
\newblock {\em Hilbert transforms vol. I--II}.
\newblock Cambridge University Press, 2009.

\bibitem{Morse+Feshbach1953b}
P.~M. Morse and H.~Feshbach.
\newblock {\em Methods of Theoretical Physics}, volume~2.
\newblock McGraw-Hill, New York, 1953.

\bibitem{Bohren+Huffman1983}
C.~F. Bohren and D.~R. Huffman.
\newblock {\em Absorption and Scattering of Light by Small Particles}.
\newblock John Wiley \& Sons, New York, 1983.

\bibitem{Osborn1945}
J.~A. Osborn.
\newblock Demagnetizing factors of the general ellipsoid.
\newblock {\em Phys. Rev.}, 67:351--357, 1945.

\bibitem{Zwillinger2003}
D.~Zwillinger.
\newblock {\em {CRC} Standard Mathematical Tables and Formulae}.
\newblock Chapman \& Hall/{CRC} Press {LLC}, 31st edition, 2003.

\bibitem{Trepte+etal2016}
C.~J.~C. Trepte, C.~R. Phillips, J.~Sol\`{a}, A.~Adler, S.~A. Haas, M.~Rapin,
  S.~H. B\"{o}hm, and D.~A. Reuter.
\newblock Electrical impedance tomography {(EIT)} for quantification of
  pulmonary edema in acute lung injury.
\newblock {\em Critical Care}, 2(18):1--9, 2016.

\end{thebibliography}

\end{document}